# Agile spectral tuning of high order harmonics by inteference of two driving pulses

**VITTORIA SCHUSTER,**[1,*] **VINZENZ HILBERT,**[1] **ROBERT KLAS,**[1,2] **JAN ROTHHARDT,**[1,2,3] **JENS LIMPERT**[1,2,3]

[1]*Institute of Applied Physics, Friedrich-Schiller University Jena, D-07745 Jena, Germany*
[2]*Helmholtz-Institute Jena, D-07743 Jena, Germany*
[3]*Fraunhofer Institute of Applied Optics and Precision Engineering, D-07745 Jena, Germany*
*\*Corresponding author: vittoria.schuster@uni-jena.de*

**In this work the experimental realization of a tunable high photon flux extreme ultraviolet light source is presented. This is enabled by high harmonic generation of two temporally delayed driving pulses, resulting in a tuning range of 0.8 eV at the 19[th] harmonic at 22.8 eV. The implemented approach allows for fast tuning of the spectrum, is highly flexible and is scalable towards full spectral coverage at higher photon energies.**

Over the past three decades high harmonic generation (HHG) of ultrashort driving laser pulses has emerged as a complimentary technique to large-scale facilities for the generation of extreme ultraviolet (XUV) light. The broad spectral coverage of the harmonic comb, coherence and ultrashort pulse duration (fs - as) [1], achievable with HHG have enabled a large variety of table-top applications. These range from imaging experiments with extreme spatial resolution [2] to XUV spectroscopy, which gives access to the electronic structure of atoms, ions, molecules and solids [3–12]. Often, selecting one narrow bandwidth single harmonic line with a high photon flux is beneficial for such experiments. However, the discrete orders of the harmonic comb, which are odd multiples of the driving laser frequency, can constitute a severe limitation for these experiments. Tunable HHG sources can increase the spectral coverage while maintaining a narrow linewidth and high flux, thus enabling to target specific resonances regardless of their spectral position. This holds immense potential for aforementioned spectroscopic techniques and has been employed for the generation of quasi-supercontinua [13]. Furthermore, tunable XUV sources are of great interest for lens-less XUV multispectral imaging. Here, scanning the light source over a characteristic resonance of the sample provides access to spectral information [14], which otherwise needs to be retrieved from broadband measurements with larger computational [15–18], or experimental effort [19–21].

A large variety of spectrally tunable HHG sources have been demonstrated, often relying on manipulation of the driving laser field. Most intuitively, spectrally tunable infrared sources have been employed, either by high-order frequency mixing the weak but tunable output of OPG with a fixed-frequency high pulse energy laser inside the gas target [22,23], or by directly driving the HHG process with a tunable optical parametric amplifier (OPA) output [13,24,25]. Other approaches exploit the HHG spectrum's dependence on the driving laser intensity through blue-shifting [26], on the pulse chirp [27], or both [28]. Furthermore, the emergence of devices for tailored modification of the spectral phase and amplitude of the driving laser has led to general flexibility in the custom control of HHG spectra [29–31].

Despite the numerous implementations, demand for an efficient tunable HHG source with low complexity, both with regards to the setup and the optimization parameters, remains unabated. Oldal *et al.*[32] recently presented theoretical considerations and simulation results for a tunable high-harmonic generation source that relies solely on the time delay between two identical driving pulses of fixed wavelength to shift the HHG spectrum. An implementation of this approach at existing HHG sources is possible without modification to the driving laser, promising an accessible route to tuning narrowband XUV spectra. In this work we present the first experimental demonstration of a tunable HHG source following the two-pulse method proposed in Ref. [32].

The approach is based on shifting the central photon energy $E_{\text{fund}}$ of the fundamental driving pulse, which translates to a change in photon energy of the generated XUV radiation. The central energy of the fundamental is tuned solely by exploiting the spectral interference of two identical delayed pulses of constant spectrum. Fig. 1 shows spectra of a thus composed double pulse for several delays. For a given path delay d, different wavelength components $\lambda$ of the spectrum interfere differently, since the introduced phase shift is given by $\varphi = 2\pi\, d/\lambda$. The resulting spectral fringe pattern changes in position and width as the delay between the pulses is tuned. As is illustrated in Fig. 1, for increasing delay the fringe pattern is shifted from smaller to bigger wavelengths. For sufficiently wide fringes, here corresponding to delays around 13 times the driving wavelength, broad portions of the spectrum interfere destructively. These components thus do not contribute to the resulting pulse and its mean wavelength. For the delay range depicted in Fig. 1, this results in a continuous increase of the central wavelength between the red and the yellow curve. As a consequence, the central energy of the double pulse's spectrum can be shifted by choosing an appropriate delay.

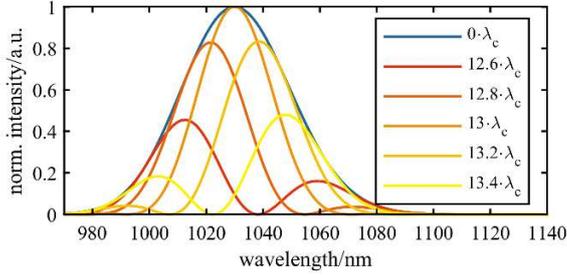

Fig. 1. Simulated spectra of a double pulse for different delays between the individual pulses (Fourier-limited duration: 35 fs), given in multiples of the central wavelength $\lambda_C$, 1030 nm. Due to the constructive and destructive interference of different parts of the spectrum, the central wavelength is shifted with changing delay. The blue curve depicts the spectrum for zero path delay, which is analogous to the spectrum of the individual pulses. The spectra are retrieved through Fourier transformation of the temporal double pulses.

In good approximation, the photon energy of the high-order harmonic of order $H$ driven by a fundamental field with central photon energy $E_{\text{fund}}$ can be expressed as $H \cdot E_{\text{fund}}$. Thus, by changing the central energy of the fundamental by $\Delta E_{\text{fund}}$, the central energy of the harmonic changes by $\Delta E_H$:

$$E_H = H \cdot E_{\text{fund}} \Rightarrow \Delta E_H = H \cdot \Delta E_{\text{fund}}$$

As can be seen in Fig. 1, the different interference conditions that are created by changing the delay do not only influence the double pulse spectrum but also its intensity. Consequently, the generating conditions of the XUV light vary for different delays and some level of adjustment is necessary to ensure efficient high-harmonic generation for a large delay range. While higher intensities are favorable for the single-atom response of the generating medium ($\propto I^9$ [33]), they do not necessarily lead to a larger XUV flux due to phase-matching effects. Especially the increase of the ionization fraction above the critical ionization, beyond which true phase-matched HHG is not possible [34], limits the driving laser intensity for a given pulse duration. While phase matching is influenced by many parameters such as focusing geometry, gas pressure and target position, these are in the following kept constant within one measurement for the sake of experimental simplicity and comparability. The only phase matching adjustment between individual measurements is the size of an iris aperture before the focusing lens. Simply speaking, further closing the aperture reduces the intensity in the generating medium, thus favoring delays for which the interference between the driving pulses is more constructive.

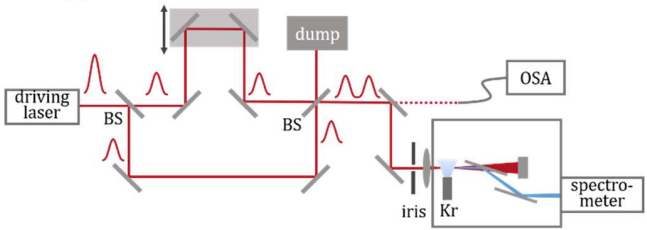

Fig. 2. Schematic of the setup for tunable high-harmonic generation. BS: beam splitter; OSA: optical spectrum analyzer

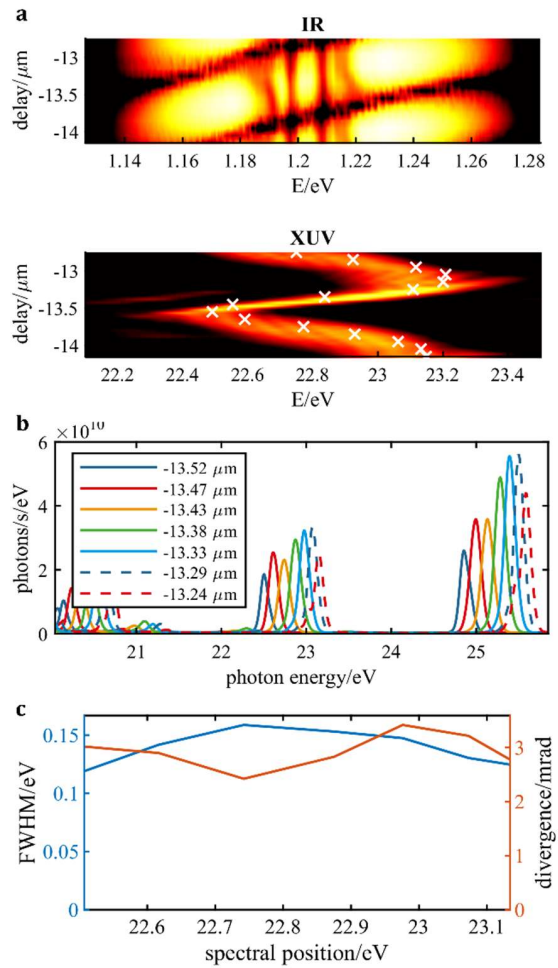

Fig. 3. Tuning of H19 by changing solely the delay between the driving pulse pair. a: Logarithmic spectra of the driving double pulse (top) and H19 (bottom) over a delay range of more than one driving laser cycle. White crosses in the bottom plot represent the central photon energy of the driving double pulse multiplied with the harmonic order 19. b: XUV spectra of H17-H21 between delays of -13.24 µm and -13.52 µm. c: bandwidth and divergence of the harmonic lines depicted in b.

In this experiment, the driving double pulse is composed of two time-delayed pulses of equal energy and duration by a Michelson-type interferometer. This approach allows for continuous tuning over a large delay range and only utilizes elemental optical components, which support scaling to high powers and broad bandwidths. Furthermore, minimizing the nonlinear effects occurring in solid optical materials results in a highly flexible setup that is directly implementable at the output of systems with short pulses and high average powers.

A schematic of the setup is depicted in Fig. 2. The output of a commercial pulsed 35 fs 200 kHz Yb fiber amplifier system with a hollow core fiber compression stage is split into two partial beams with pulse energies of 37 µJ by an intensity beamsplitter. A high-precision translation stage in one of the interferometer arms allows for continuous control of the relative delay between the two partial beams, which are superimposed again at a second beamsplitter. The spectrum of the resulting combined beam is monitored through a mirror leak with an optical spectrum analyzer. The beam itself passes through an iris aperture before entering a vacuum chamber in which it is focused into a Krypton gas jet with a backing pressure of 5-5.5 bar for HHG. The generated XUV light is separated from the driving infrared by two grazing-incidence plates [35] and a 200 nm Al filter and consequently analyzed with a flat-field grating spectrometer. The HHG process was arbitrarily optimized for the 19th harmonic (H19) and the tuning ranges of H17 and H21 partially lie within the analyzed spectral range.

The spectra shown in Fig. 3a and Fig. 3b are acquired while continuously decreasing the delay between the two pulses over slightly more than one driving laser cycle at a speed of 0.1 μm/s. The infrared spectra are consequently post-processed to remove uncompressed components that do not contribute to HHG It can be seen in Fig. 3a that the centers of both the driving and the XUV spectrum shift from smaller to bigger photon energies over the constructively interfering part of the cycle. Since the depicted delay range is 13 driving laser wavelengths away from zero path delay, the driving pulses do not completely cancel each other out at the point of destructive interference. Therefore, a weaker and broader harmonic line shifting back to smaller photon energies is produced in this delay range.

The scanning speed and the spacing between the measurements in Fig. 3 are solely limited by the acquisition speed of the spectrometers. Without any further adjustment of the HHG parameters or iris aperture, the spectrum of the 19th harmonic is tuned over 0.6 eV in less than three seconds, with just 20% relative standard deviation of the peak flux. Moreover, both the spectral bandwidth and the divergence, which are depicted in Fig. 3c, maintain a relative standard deviation of 11% over this tuning range. This constitutes a remarkable ease and speed of tuning the spectral position of the high harmonic while maintaining notable constancy of the further parameters.

While this approach offers the fastest tuning speed, the spectral tuning range over which a constant XUV flux is achieved can be extended with an additional adjustment of the diameter of the iris aperture in the driving beam. The resulting measurements are depicted in Fig. 4. The central energy of H19 is tuned over a range of $\Delta E_{\text{XUV,max}} = 0.8$ eV while keeping the relative standard deviations of both peak flux and bandwidth at 12% and of the divergence at 14 %. The achieved tuning range corresponds to 3.5% of the photon energy of H19 and 33% of the spectral spacing between two harmonic lines, 2.4 eV. Please note that these measurements were acquired without any changes to the backing pressure and HHG jet position.

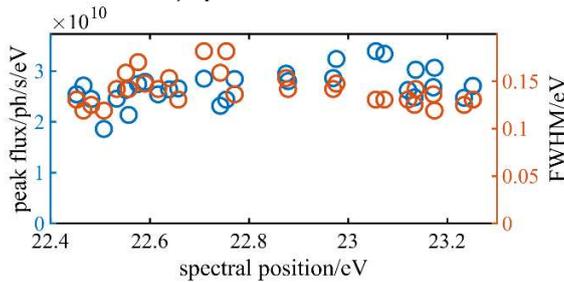

Fig. 4. Peak flux and bandwidth of H19 for different delays and iris aperture diameters.

In the following, we will discuss paths to increase the tuning range towards complete spectral coverage. The maximum achievable spectral tuning range of the XUV light depends mainly on two parameters: the harmonic order and the spectral bandwidth of the driving laser.

To illustrate the influence of the harmonic order, the spectral position of multiple harmonics is evaluated for two delays. Those two delays, 7.9 μm and 8.3 μm, are chosen such that H17, H19 and H21 are all within the spectral range covered by the detector chip and furthermore the second diffraction orders of H35 and H37 are clearly visible. The diffraction angle for the second order is equivalent to the first order of light having half its photon energy. Consequently in Fig. 5a H35 and H37 appear between H17 and H19, although very weakly due to the reduced diffraction efficiency of the grating. The resulting measured tuning ranges of the different harmonic orders are plotted in Fig. 5b. As derived above, the relationship between the tuning range of a harmonic $\Delta E_H$ and its order $H$ should be linear. The slope of the resulting linear fit matches the tuning range of the fundamental central energy, 0.03 eV per harmonic order.

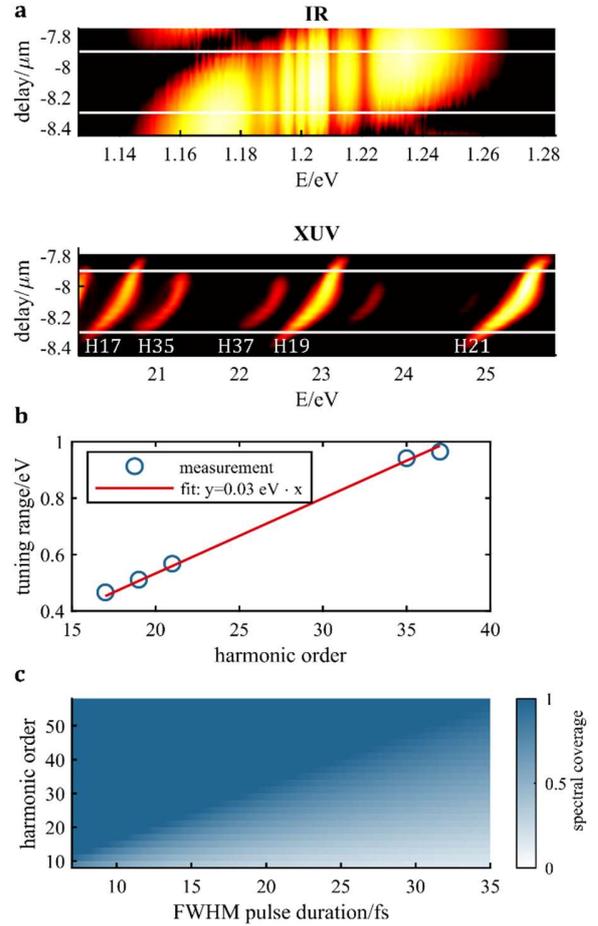

Fig. 5. Scaling of the tuning range. a: Logarithmic spectra of the driving double pulse (top) and resulting XUV spectra (bottom). The white lines mark the delays evaluated in b. Only 45% of the available pulse energy was utilized for this measurement. b: Evaluation of the spectral tuning rages of the peaks of H17-21, H35 and H37 between the delays of -8.3 μm and -7.9 μm. The red line shows a linear fit to the data, in good agreement with the expected slope, 0.03 eV/HHG order. c: Predicted dependence of the XUV spectral coverage on the Fourier-limited duration of the individual pulses at 1030 nm and the harmonic order.

Assuming the maximum tuning range of the fundamental $\Delta E_{\text{fund,max}}$ to be proportional to the full width at half maximum frequency bandwidth $BW$, the following relation can be deduced for Fourier limited individual pulses of duration $\tau_{\text{fund}}$:

$$\Delta E_{\text{fund,max}} \propto BW \propto 1/\tau_{\text{fund}}$$
$$\Rightarrow \Delta E_{\text{XUV,max}} = H \cdot \Delta E_{\text{fund,max}} \propto H/\tau_{\text{fund}}$$

An estimation of the achievable spectral coverage based on this relation and the results depicted in Fig. 4 is shown in Fig. 5c. The dark blue area marks parameter combinations with complete spectral coverage. For the implemented pulse duration of 35 fs and 1.03 μm driving laser wavelength, gap-free spectral tuning is expected for harmonic orders larger than 55 (photon energy > 66 eV). This can be reduced by using shorter pulses or larger wavelengths $\lambda_{fund}$. For a given pulse duration, the minimal harmonic order for continuous tuning $H_{min}$ scales with $1/\lambda_{fund}$, roughly halving $H_{min}$ for 2 μm drivers, and doubling it for 0.5 μm in relation to the value derived from Fig. 5c.

One advantage of the interferometric tuning approach used in this experiment is that no part of the driving laser power is lost a priori. If no change of the central energy is needed, the relative delay can be set to zero. In this case the interference of the two pulses is fully constructive and the full power of the fundamental laser used for HHG. Indeed, similar XUV spectra and efficiencies have been achieved by bypassing the interferometer, thus using a single driving pulse with the same overall pulse energy. Switching between using the full IR spectrum or parts of it is possible by changing just one parameter: the delay. This underlines the efficiency and versatility of the interferometric approach.

In conclusion, we demonstrate an efficient, quickly tunable HHG source driven by time-delayed double-pulses. The narrowband 19th harmonic of a 1030 nm driving laser is shown to be tunable from 22.3 eV to 23.1 eV for the laser parameters utilized in this work. The development potential to full spectral coverage is discussed. This work opens numerous possibilities to tailor this tuning technique to different requirements. The employed Michelson-type interferometer can easily be modified to support a wide range of driving laser parameters and does not rely on prior adjustment of the pulse properties. Due to these qualities it is outstandingly suitable for commercial driving lasers or user facilities with limited beam time. Since only highly reflective mirrors and a beam splitter are required, it is easily power scalable up to the kW regime and to broadband driving lasers. Furthermore, the high tuning speed of this approach could be further increased by realizing the delay e.g. with a piezo driven mirror, making it attractive for the generation of quasi-supercontinua [13], thus benefitting applications such as XUV Coherence Tomography [36]. By implementing an electro-optical modulator even pulse-to pulse tuning appears feasible.